\documentclass[fleqn,usenatbib]{mnras}

\usepackage{newtxtext,newtxmath}

\usepackage[T1]{fontenc}

\DeclareRobustCommand{\VAN}[3]{#2}
\let\VANthebibliography\thebibliography
\def\thebibliography{\DeclareRobustCommand{\VAN}[3]{##3}\VANthebibliography}

\usepackage{graphicx}	
\usepackage{amsmath}	
\usepackage{amssymb}	

\title[The N$_2$ Production Rate in Comet C/2016 R2]{The N$_2$ Production Rate in Comet C/2016 R2 (PanSTARRS)}

\author[S. E. Anderson et al.]{
S. E. Anderson,$^{1}$\thanks{E-mail: sarah.anderson@univ-fcomte.fr}
P. Rousselot,$^{1}$
B. Noyelles,$^{1}$
C. Opitom,$^{2}$
E. Jehin,$^{3}$
D. Hutsem\'ekers,$^{3}$
and J. Manfroid$^{3}$
\\
$^{1}$ Institut UTINAM UMR 6213 / CNRS, Univ. Bourgogne Franche-Comté, OSU THETA, BP 1615,
25010 Besançon Cedex, France\\
$^{2}$ Institute for Astronomy, Univ. of Edinburgh, Royal Observatory, Edinburgh EH9 3HJ, UK\\
$^{3}$ STAR Institute, Universit\'e de Li\`ege, All\'ee du 6 Ao\^ut 19c, 4000 Li\`ege, Belgium
}

\date{Accepted XXX. Received YYY; in original form ZZZ}

\pubyear{2015}

\begin{document}
\label{firstpage}
\pagerange{\pageref{firstpage}--\pageref{lastpage}}
\maketitle

\begin{abstract}
Observations of comet C/2016 R2 (PanSTARRS) have revealed exceptionally bright emission bands of 
N$_2^+$, the strongest ever observed in a comet spectrum. Alternatively, it appears to be poor in CN 
compared to other comets, and remarkably depleted in H$_2$O. Here we quantify the N$_2$ 
production rate from N$_2^+$ emission lines using the Haser model.
We derived effective parent and daughter scalelengths for N$_2$ producing N$_2^+$. This is the first
direct measurement of such parameters. Using a revised fluorescence 
efficiency for N$_2^+$, the resulting production rate of molecular nitrogen is inferred to be 
Q(N$_2$) $\sim$ 1 × 10$^{28}$ molecules.s$^{-1}$ on average for 11, 12 and 13 Feb. 2018, the 
highest for any known comet. Based on a CO production rate of Q(CO) $\sim$ 1.1 × 10$^{29}$ 
molecules.s$^{-1}$, we find Q(N$_2$)/Q(CO)$\sim$0.09, 
which is consistent with the N$_2^+$/CO$^+$ ratio derived from the observed intensities of
N$_2^+$ and CO$^+$ emission lines. We also measure significant variations in this production rate 
between our three observing nights, with Q(N$_2$) varying by plus or minus 20\% according
to the average value.
\end{abstract}




\begin{table*}[b]
  \caption{Observing circumstances of comet C/2016 R2 (PanSTARRS) with VLT/UVES. Table from \citet{Opitom2019}.} 
  \centering 
    \begin{tabular}{cccccccc}
    \hline
    Date  & $r_h$ (au) & $ \dot{r}_h$ (km.s$^{-1}$) & $\Delta$ (AU) & $ \dot{\Delta}$ (km.s$^{-1}$) & Exposure time (s) & UVES Setup & UVES slit\\
     \hline
2018-02-11T00:27:07.326 & 2.76 & -6.09  & 2.40 & 19.7 & 4800 & DIC1-390+580 & 0.44''$\times$8'' - 0.44''$\times$12'' \\

2018-02-13T00:46:23.196 & 2.76 & -5.97  & 2.43 & 19.9 & 4800 & DIC1-390+580 & 0.44''$\times$8'' - 0.44''$\times$12'' \\ 

2018-02-14T00:47:40.759 & 2.75 & -5.91  & 2.44 & 20.1 & 4800 & DIC1-390+580 & 0.44''$\times$8'' - 0.44''$\times$12'' \\

    \hline
    \end{tabular}
\label{tab:obs}
  \end{table*}

\section{Introduction}

Comets are among the most pristine relics of the formation of the Solar System, having agglomerated from different icy grains and dust particles leftover from the planetary formation process and undergone little alteration since. As comets approach the Sun, the sublimation of their ices creates a large gaseous coma, which, along with the interaction with solar radiation, leads to various spectroscopic emissions. This allows us to investigate the composition of cometary ices and determine the physico-chemical conditions of their formation, thus providing further insight to the nature of the early Solar System at the time and place these comets formed.

The long period comet C/2016 R2 (PanSTARRS), displayed atypical coma morphology in optical images since its passage near perihelion ($\sim$ 3.0 au) in December 2017, emitting strongly in blue optical wavelengths due to ion emission dominating in the coma. Radio observations revealed that it was a CO-rich comet \citep{Wierzchos2018} and strongly depleted in water, with an H$_2$O/CO ratio of 0.0032 \citep{McKay2019} with an upper limit of $<$ 0.1 \citep{Biver2018}. The spectrum was dominated by bands of CO$^+$ as well as N$_2^+$, the latter of which was rarely seen in such abundance in comets \citep{Cochran2018,Opitom2019}. It was also found to be both CN-weak and dust-poor, with an $Af\rho$ of 500 cm \citep{Opitom2019}. This CO-rich and water-poor composition, along with none of the usual neutrals seen in most cometary spectra, makes C/2016 R2 a unique and intriguing specimen. 

The apparent N$_2$ deficiency in comets has long been a matter of great debate. Despite Pluto and Triton -- which also formed in the outer Solar System -- both exhibiting a N$_2$-rich surface \citep{Cruikshank1993, Owen1993, Quirico1999, Merlin2018}, very few ground-based 
facilities have ever observed N$_2^+$ in cometary spectra.  This mainly concerns the following 
comets: C/1908 R1 (Morehouse) \citep{Delabaumepluvinel1911}, C/1961 R1 (Humason) 
\citep{Greenstein1962}, 1P/Halley
\citep{Wyckoff1989,Lutz1993}, C/1987 P1 (Bradfield) \citep{Lutz1993}, 29P/Schwassmann-Wachmann~1 
\citep{Korsun2008,Ivanova2016,Ivanova2018}, and C/2002 VQ94 (LINEAR) \citep{Korsun2008,Korsun2014}, 
and potentially C/2001 Q4 (NEAT) \citet{Feldman2015}. Only C/2002 VQ94 presents a N$_2$/CO 
ratio as high as C/2016 R2.

The detection of N$_2$ in comets using spectroscopic methods has been challenging: as a diatomic, symmetrical molecule, N$_2$ has no permanent dipole moment, which results in the absence of pure rotational transitions. It emits no radiation at millimeter-wavelengths, making the molecule invisible to observations in that range. The electronic transitions are visible in the UV through instruments placed outside of our atmosphere. The presence of N$_2$ in comet 67P/ Churyumov-Gerasimenko was not detected through any transition but using the ROSINA mass spectrometer in-situ measurements \citep{Rubin2015}. However, its daughter-species N$_2^+$ is detectable in the optical range through the bands of the first negative group (B$^2\Sigma_u^+$ 
- X$^2\Sigma_g^+$) with the (0,0) bandhead located at 3914~\AA. Not only have this ion's spectral lines been observed in C/2016 R2, they are also the brightest ever seen in a cometary spectrum \citep{Cochran2018}. The quantity of N$_2$ present is thus of significant importance. By measuring the band intensity of the observed N$_2^+$ in C/2016 R2's spectrum, assuming that solar resonance fluorescence is the only excitation source, the observed emission fluxes have been used to calculate ionic ratios of N$_2^+$/CO$^+$ in the coma between $0.06 \pm 0.01$ \citep{Cochran2018,Opitom2019} and $0.08 \pm 0.01$ \citep{Biver2018}. This would be the same 
ratio for N$_2$/CO since ionization efficiencies of N$_2$ and CO are similar ($\beta_{i,N_2} = 3.52 
\times 10^{-7}s^{-1}$ and $\beta_{i,CO} = 3.80 \times 10^{-7}s^{-1}$ at 1 au for quiet Sun 
\citep{Huebner1992}) as long as they are fully photodissociated in the coma. This is larger than the best measurement in comet 67P, with a N$_2$/CO ratio of $\sim 2.87 \times 
10^{-2}$\citep{Rubin2020}, though these measurements were obtained much closer to the nucleus.

\begin{figure*}
\centering
	\includegraphics[width=17.8cm]{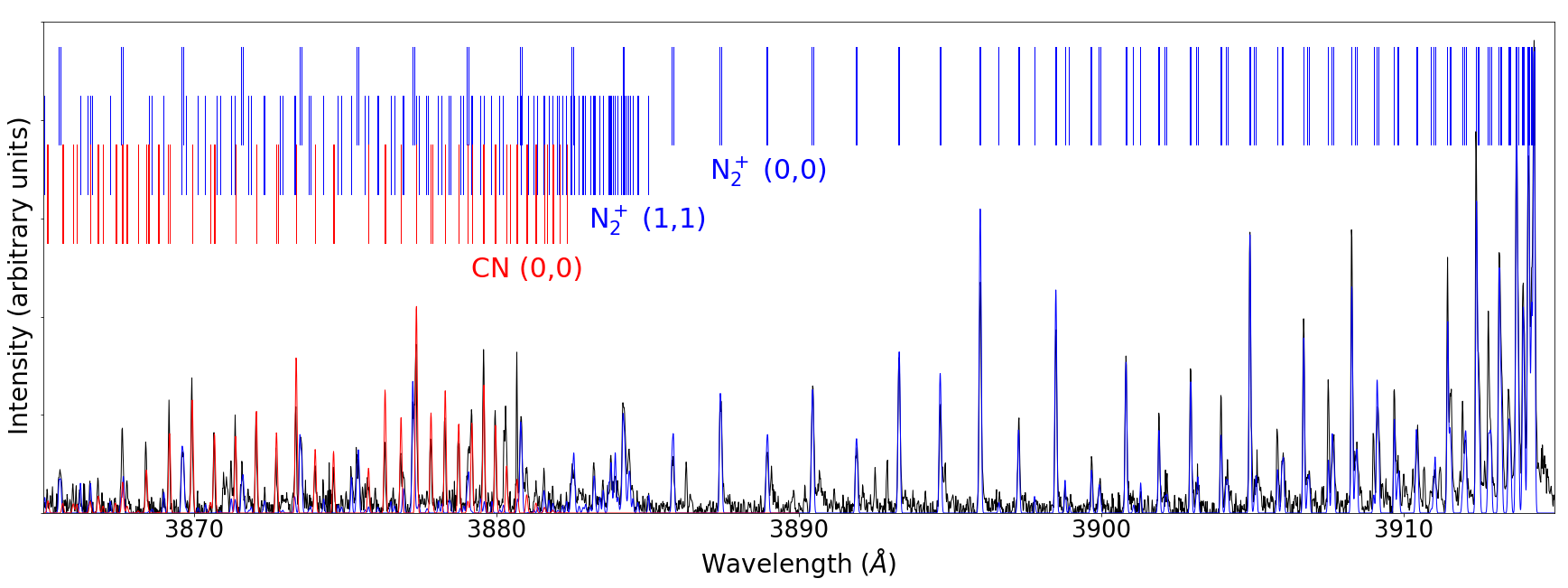}
    \caption{Observational spectrum of C/2016 R2 (black) obtained at the edges of the slit ($\sim6500$km), overlapped
    with a CN model (red) and a N$_2^+$ model (blue) from \citet{Rousselot2022}. The $y$ axis (intensity scale) has arbitrary units.}
    \label{fig:Model}
\end{figure*}

The origin of the unusual composition of C/2016~R2 is controversial. It may be a fragment of a 
differentiated object as suggested by \citet{Biver2018}, similar to what was suggested for the CO-rich
Interstellar comet 2I/Borisov \citep{Cordiner2020}. There were also suggestions that the high CO 
content in C/2016 R2 might be due to it being formed just beyond the CO snowline in the protosolar
nebula \citep{Eistrup2019,Mousis2021}. Understanding the exact composition of this comet is essential 
to determine the likelihood of each formation scenario, adding additional insight into the 
composition of the early Solar System.

Using the N$_2^+$  emission lines identified in the spectrum of C/2016 R2, we can then determine how 
much N$_2^+$ is being produced through the use of the Haser model. This model relates the intensity 
of the observation to the number of molecules responsible for this emission. We can also determine 
scalelengths related to the lifetime of the N$_2$ molecules and N$_2^+$ ions in the coma of
C/2016 R2, which can be used for future comets presenting high quantities of N$_2^+$. This approach allows the determination of how much N$_2^+$, thus N$_2$, is being produced by the active surface of the 
comet's nucleus.

Understanding the presence - or lack of - N$_2$ in comets allows us to investigate key features in the timeline of planetesimal formation. An object such as C/2016 R2 is a unique opportunity to set a baseline for identifying and exploring N$_2$ in cometary spectra. We begin by describing the method of obtaining the UVES (the high resolution spectrograph of the European Southern Observatory Very Large Telescope) observations of C/2016 R2. Then, we outline the Haser model and how we apply it to these observations in the \textsc{Methods} section (Sec. \ref{sec:methods}). We start by applying the model to the molecule of CN, as it is well known in the literature and provides a reliable benchmark for which to test the quality of our observations and application of our methods. Finally, we describe how N$_2^+$ is identified in the spectra of cometary coma and how we constrain its properties, before calculating its production rate. We then discuss the implications of these results and conclude as to how this can be applied to future N$_2$-rich comets.

\section{Observations}\label{sec:obs}

The observations of C/2016 R2 which were used in our analyses were obtained on 2018 February 11, 13, and 14 (one single exposure of 4800~s of integration per night) with the Ultraviolet-Visual Echelle Spectrograph (UVES) \citep{Dekker2000} mounted on the 8.2 m UT2 telescope of the European Southern Observatory Very Large Telescope (ESO VLT). All observations were made when C/2016 R2 was near its perihelion distance of 2.6~au, at 2.76~and 2.75~au. We used a 0.44'' wide slit, providing a resolving power of R$\sim$80,000. The slit length was 8'', corresponding to about 14,500~km at the distance of the comet (geocentric distance varying from 2.40 to 2.44~au). This is summarized in Table \ref{tab:obs}. No N$_2^+$ atmospheric lines were detected in the high resolution UVES spectra.

The ESO UVES pipeline along with custom routines was used to perform the data reduction; extraction; 
cosmic rays removal; and correct for the Doppler shift due to the relative velocity of the comet with 
respect to the Earth. Both the archived master response curve or the response curve determined from a 
standard star observed close to the science spectrum were used to calibrate the spectra in absolute 
flux using either with no significant differences. More details can be found in the UVES ESO pipeline 
manual. This data processing produced a 2D spectrum for each night of 11, 13, and 14 Feb, calibrated in 
wavelength and absolute flux units. The spatial extent reached 30 rows each corresponding to 
0.25~arcsec on the slit with different cometocentric distances centred on the nucleus. A full 
description of the observations and further data reduction can be found in \citet{Opitom2019}.

\section{Methods}\label{sec:methods}

\begin{figure}
\centering
	\includegraphics[width=\columnwidth]{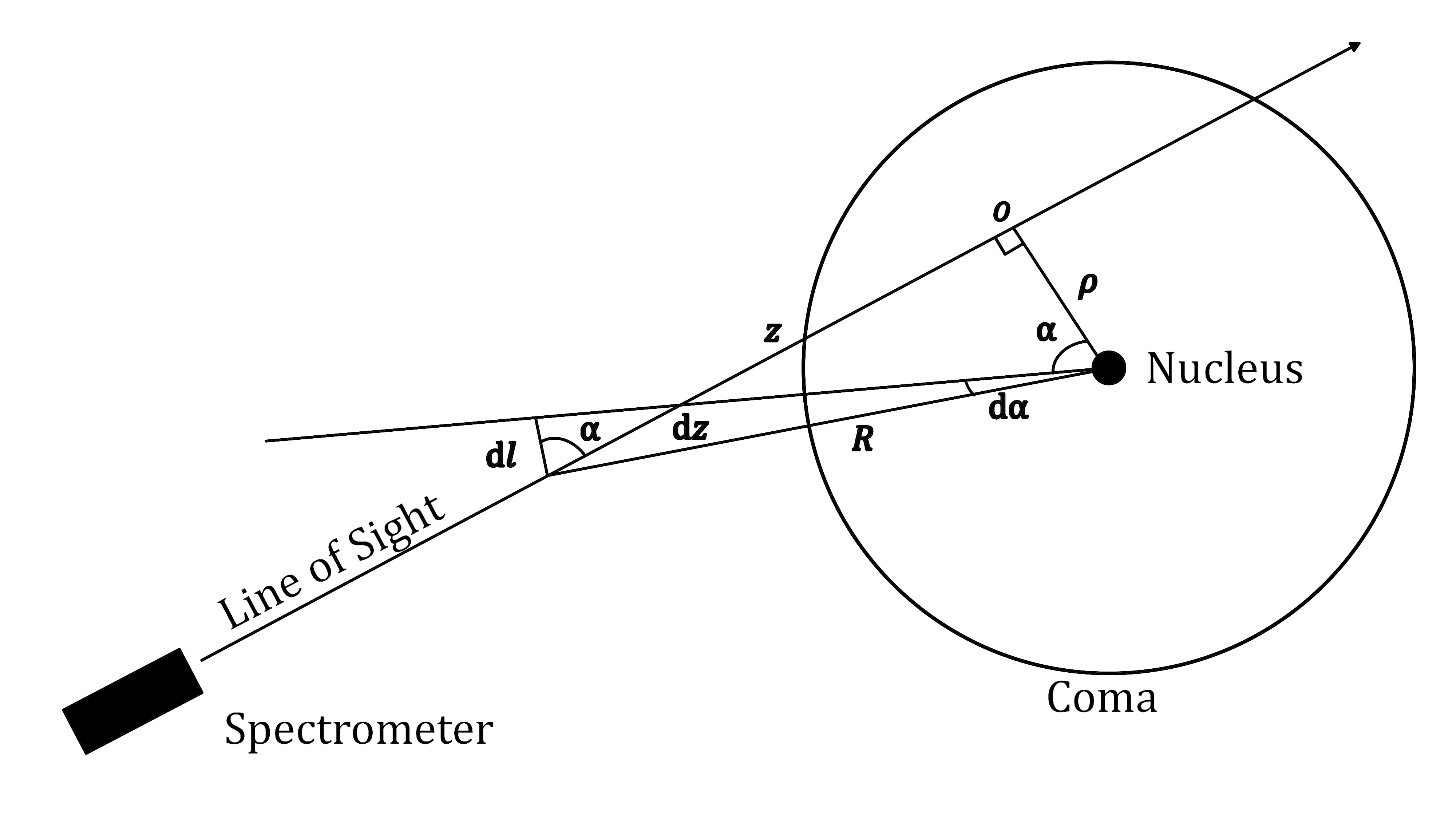}
    \caption{Line of sight and variables used for the integration of the Haser profile. The origin of the coordinates is the location at which $R$ is minimum.}
    \label{fig:Haser}
\end{figure}

A usual method for estimating the absolute gas production rates in a cometary coma is to fit the observed flux to a Haser profile \citep{Haser1957}. This gives an analytical solution to the column density of parent and daughter species in the coma along the line of sight. Assuming the medium optically thin, the intensity of the coma observed at a projected distance 
$\rho$ from the nucleus would be proportional to the number of molecules along the line of sight that 
produce it, and we can infer how much of the considered species is present in the coma at that 
specific cometocentric projected distance. These parameters are displayed in Fig. \ref{fig:Haser}.

The parent molecules are disintegrated by photodissociation by UV photons following the law $n = n_0 \mathrm{e}^{-t/\tau_p}$, with $n_0$ the number of parent molecules at $t$ = 0 and $\tau_p$ the 
average lifetime of a parent molecule. The daughter-species produced from the photodissociation of 
the parent-molecules are seen to be ejected radially with a velocity of $v_d$ from the nucleus. 
In the case of $N_2$ it is an ionization process but it obeys this hypothesis in the Haser model. 

The photodissociation (or photoionization) lifetime, $\tau_p$, is an important factor in 
the behaviour of the parent and daughter species. We use the relationship $ l_{p,d} =\tau_{p,d} v_{p,d}$ which determines the scalelength of the parent or daughter species, respectively. If we 
call $Q = n(R_0)\cdot 4\pi R_0^2 v_p$ the production rate (in molecules.$s^{-1}$) of a given parent species and $n(R)$ the volume density of the parent species at distance $R$ from the nucleus (of radius $R_0$), we then integrate over the line of sight:

\begin{equation}
N_{\text{parent}}(\rho) = \mathrm{e}^\frac{R_0}{l_p}\cdot \frac{Q}{2\pi v_p \rho} \int_{0}^{+\pi/2}\mathrm{e}^\frac{-\rho}{l_p \cos{\alpha}}\mathrm{d}\alpha
\end{equation}

\noindent where $N_{\text{parent}}(\rho)$ is the column density of the parent species in
molecules/cm$^2$,
and $\rho$ the  projected cometocentric distance. This provides a fair assessment of the number of
parent molecules at a given projected distance from the nucleus 
of the comet. Because N$_2^+$ is a daughter species we need to take into account the 
photoionization of its parent molecule at each successive distance in order to determine the 
quantity of daughter species being produced by successive disintegrations of the parent molecule. 

\begin{figure*}[b!]
	\includegraphics[width=\textwidth]{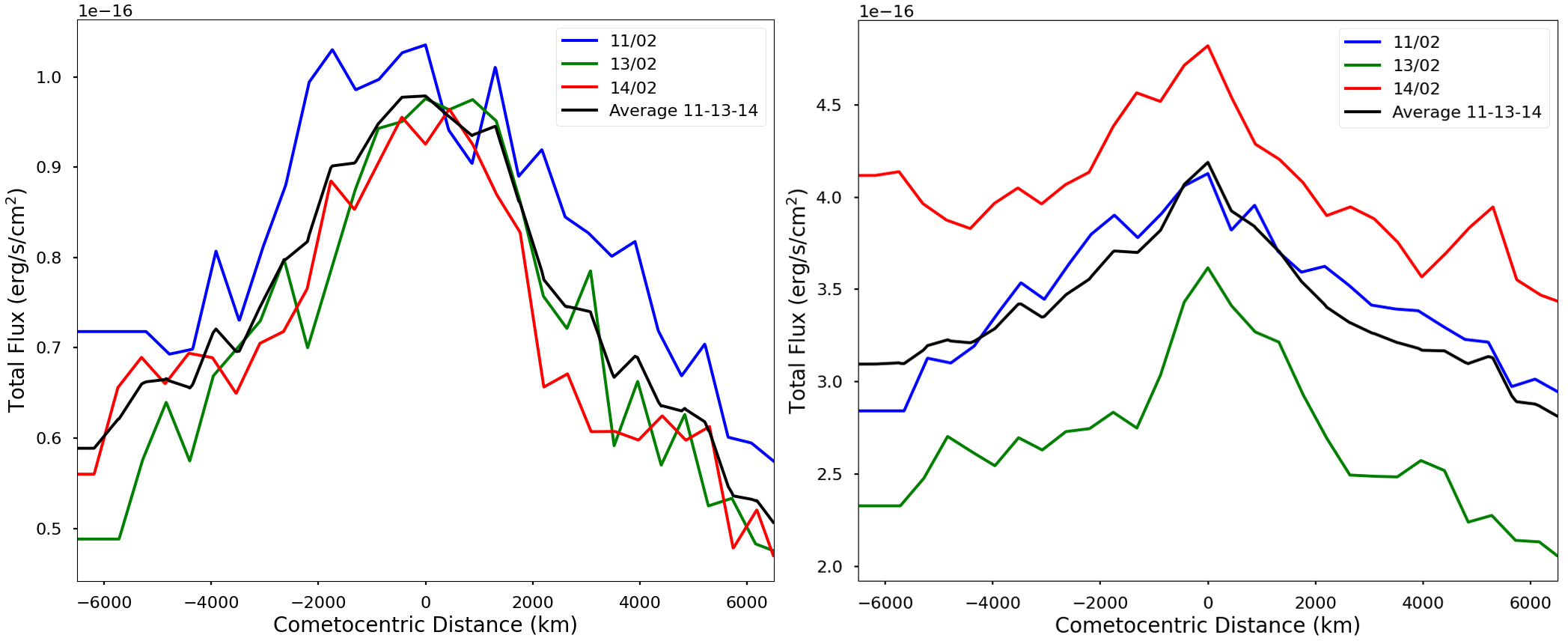}
    \caption{Intensity profiles for identified CN (left) and identified N$_2^+$ (right) for the nights of 11 (blue), 13 (green), and 14 (red) Feb over the cometocentric distances. The average value of all three nights is shown in black. The variation of the intensity of N$_2^+$ over all three nights is striking, compared to the consistency of CN. This is possibly due to the rapid variation of the ion tail, seen to change from hour to hour.}
    \label{fig:avr}
\end{figure*}

One of the limitations of the Haser model for computing the profile of a ionic species 
such as N$_2^+$ is that it considers neutral products (like OH, C$_2$, CN...) not affected 
by electric forces due to the solar wind particles (or any other ionic species present in the coma). 
This work being restricted, however, to the inner coma (projected distance inferior 
to 8000~km), this limits the effects of electric forces on the dynamics of radical species 
and the physical hypotheses of
the Haser model are still valid. \citet{Raghuram2020} computed that the dominant process for creating
N$_2^+$ ions in C/2016 R2 is the photoionization process due to solar UV photons, the ionization 
process due to electrons inside the coma representing only a few percents relative to photoionization.
The validity of using the Haser model is also confirmed by the quality of our fit (see Fig. \ref{fig:N211}) and the symetry in the observed intensity profiles (see \ref{fig:avr}).

For the daughter species N$_2^+$ we have:

\begin{equation}
N(\rho) = \frac{Q}{2\pi v_d \rho} \frac{l_d}{l_p-l_d} \int_{0}^{+\pi/2} \mathrm{e}^{\frac{1}{l_p}(R_0 - \frac{\rho}{\cos(\alpha)})} - \mathrm{e}^{\frac{1}{l_d}(R_0 - \frac{\rho}{\cos(\alpha)})} \mathrm{d}\alpha
\end{equation}

We then integrate N'($\rho$) along the slit in order to obtain $N_\text{tot}$, the total number of daughter species present in the slit for a production rate of 1~molecules.$s^{-1}$. 
We apply this value to the relationship:

\begin{equation}
N_{\text{tot}} Q = \frac{4\pi F_\text{tot}\Delta^2}{g}
\end{equation}

\noindent with $Q$ the production rate in molecules.s of the parent molecule$s^{-1}$, and $\Delta$ the 
geocentric distance of C/2016 R2 at the moment of observation. 
$F_\text{tot}$ represents the total flux observed through the slit (in erg.s$^{-1}$.cm$^{-2}$) and $g$ is the fluorescence efficiency (expressed here in erg.s$^{-1}$.molecule$^{-1}$). We can compute the production rate $Q$ with: 

\begin{equation}
Q = \frac{4\pi F_\text{tot}\Delta^2}{g N_{\text{tot}}}
\end{equation}

This formula corresponds to the simple case where all the parent species
photodissociate into the same daughter species. If this is not the case the branching ratio
(i.e. the fraction of daughter species produced compared to all the destruction processes) 
must be taken into account. The production rate given by the formula above must be divided by this
branching ratio.

To compute the production $Q$ we need, consequently, to compute the parameter $N_{\text{tot}}$ and 
$F_{\text{tot}}$ the total flux inside the slit. The first quantity implies the knowledge of the 
Haser parameters mentioned above, i.e. $v_d$, $l_p$ and $l_d$.

For $v_d$ most authors use a value of 1~km.s$^{-1}$ for this parameter,
because it is 
a reasonable estimate for comets at $\sim$1~au (based, e.g. on the results from 
space probes). \cite{Cochran1993} discuss the variability of this parameter both with 
heliocentric distance and production rate. The production rate seems to influence this parameter
only for very active comets, which is not the case for C/2016 R2. The heliocentric distance 
influences significantly the gas expansion radial velocity inside the coma. This effect 
cannot be neglected in our case because of the large heliocentric distance of C/2016 R2 at the time
of our observations (2.76~au). \cite{Cochran1993} conclude their discussion by using
the law $v_d=850 \times r_h^{-0.5}$ with $v_d$ expressed in km.s$^{-1}$. Such a law provides
$v_d=0.511$ km.s$^{-1}$. It is why we decided to use 0.5~km.s$^{-1}$, which is 
also the value used by \cite{Raghuram2021}.
It is important to keep in mind, nevertheless, that such a law is based on observations of
comets with coma dominated by water molecules. In the case of C/2016 R2 the coma is dominated 
by CO, which can have an influence on this parameter. This influence being difficult to quantify the
above mentioned value seems the best approximation that we can use. For our discussion on the
scalelengths and for a better comparison with other works, that consider scalelengths varying 
as $r_h^2$ (i.e. that consider a constant velocity with heliocentric distance and only the 
variation of lifetime with heliocentric distance) we will use the same scaling in this paper.

Because no work has ever been published so far for $l_p$ and $l_d$ values in N$_2^+$ in a cometary coma, our method consisted of computing this profile along the slit and fitting it with different sets of parameters to find the best values. The process is described in Sec. \ref{sec:CN}.These parameters were then used to compute N$_{\text{tot}}$. For the fluorescence efficiencies $g$, we used recently revised values \citep{Rousselot2022}. The true size of the radius of the nucleus is unknown; a reasonable approximation must be used. The following results are based on calculations using a 5~km radius as this is the value consistently used when applying the Haser model to comets of unknown size.

From the 2D spectra obtained with UVES, we extracted
row by row along the slit length 30 1D spectra for each of the three different observing nights. For all of them we subtracted individually 
the solar continuum scattered by the dust grains, by adjusting a solar spectrum \citep{Kurucz1984} 
convolved with the instrument response function in the region without cometary emission lines.

These 1D spectra were then used to compute the total flux $F_{tot}$ both for N$_2^+$ and CN, by summing their different emission lines (we describe this process in Sections \ref{sec:CN} and \ref{sec:n2}). Each 1D spectrum corresponding to a different cometocentric distance, it was possible to compute radial profiles and to test different fits based on a Haser profile with different sets of parameters. The seeing was taken into account for computing these profiles, by convolving the theoretical profile with a Gaussian having a FWHM similar to the seeing value recorded during the observations (i.e. 0.95'', 1.1'', and 0.94'' respectively for the nights February 11, 13, 14).

In order to properly estimate the production rate of N$_2^+$, we first measured the production rate of CN as a benchmark to ensure our methods are sound. CN has long been studied in comets, and is well referenced in the literature. We first fitted the Haser model on the CN (0,0) B$^2\Sigma^+$ - X$^2\Sigma^+$ band as a test to know if scalelengths can properly be determined from data at hand.

\section{Results and Discussion}\label{sec:res}

\subsection{CN production rate \label{sec:CN}}

\begin{figure*}
	\includegraphics[width=\textwidth]{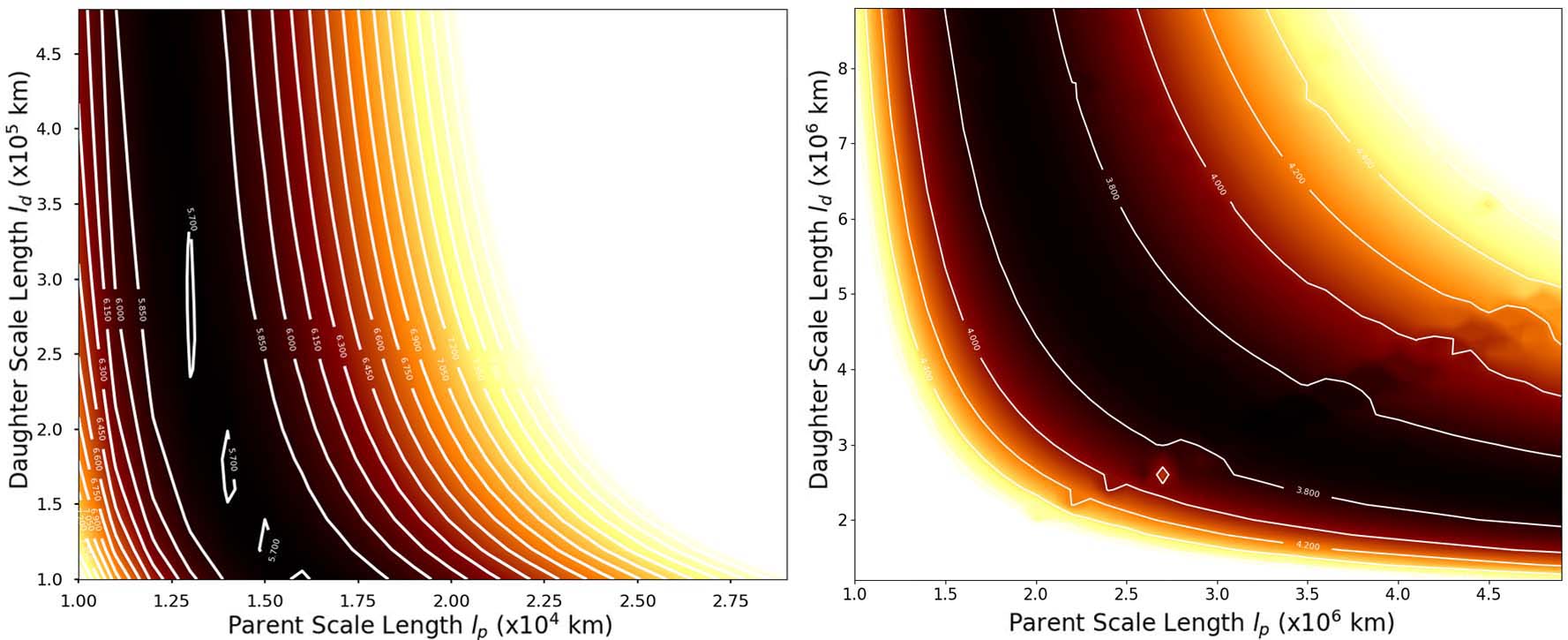}
    \caption{Confidence level contours defined from the $\chi^2$ distribution showing the range of parent and daughter scalelengths that fit the observed data for CN molecule (left) and N$_2^+$ ion (right). Values displayed are the result of the $\chi^2$ fit $\times 10^{-16}$. The darkest region indicates the best fit. Multiple pairs of parent and daughter scalelengths provide almost identical fits to the observed data. In the case of N$_2^+$, as the scalelengths are of the same magnitude, numerical errors emerge when approaching a pair of identical scalelengths as it results in a division by zero.}
    \label{fig:Chi}
\end{figure*}

\begin{figure}
	\includegraphics[width=\columnwidth]{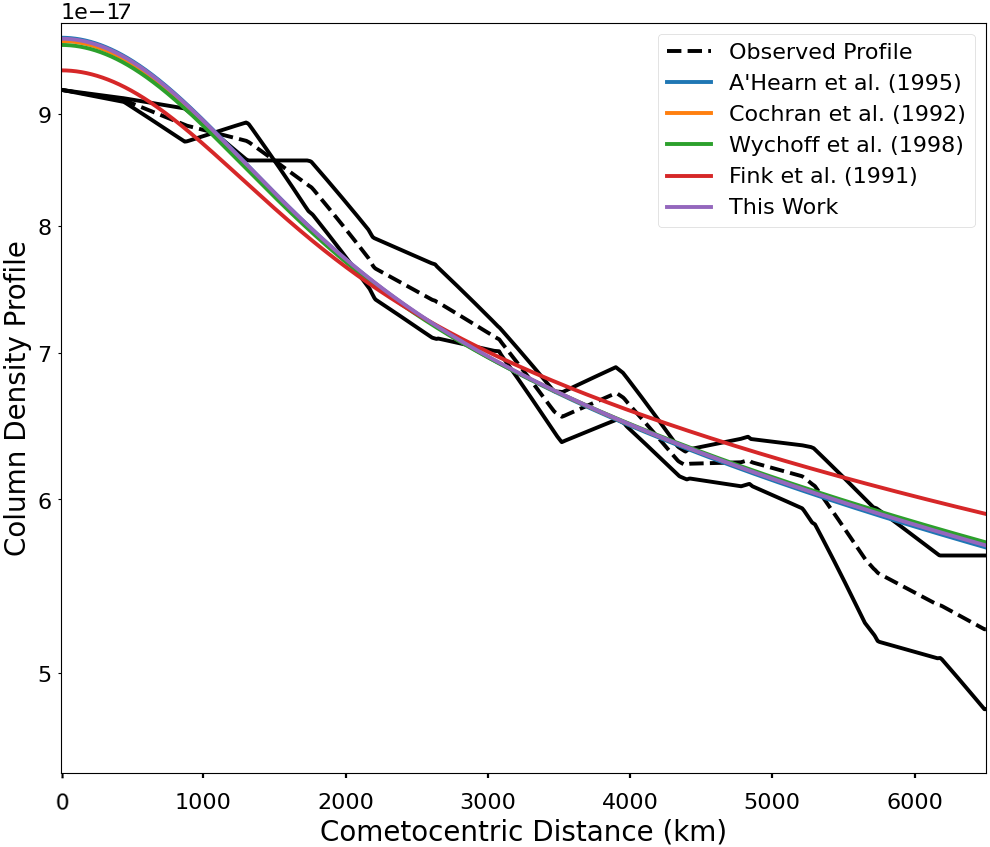}
    \caption{Our best fit of the convolved Haser model (purple) on the observed flux of CN (black) along with models from scalelengths obtained from the literature (Table \ref{tab:CN_Scale}). The observed CN flux is averaged from the observations of 11 Feb, 13 Feb, and 14 Feb (one profile for both sides of the slit from the nucleus), then averaged again over cometocentric distance (dashed line).}
    \label{fig:CN11}
\end{figure}

\begin{table}
  \caption{Parent and daughter scalelengths $l_p$ and $l_d$ at 1 au for the CN molecule found in the literature. This provides an indication of what to expect to find in Comet C/2016 R2. The values of the daughter scalelengths are not proportional to the parent scalelengths. Our fits with these scalelengths are shown in figure \ref{fig:Haser}.}
  \centering 
	\centering
    \begin{tabular}{lcc}
    \hline
    Reference  & $l_p$ [km] & $l_d$ [km]\\
     \hline
     This work & 1.3 $\times 10^4$  &   2.8 $\times 10^5$ \\ 
     \citet{AHearn1995}  & 1.3 $\times 10^4$  &   2.2 $\times 10^5$ \\
     \citet{Cochran1992} & 1.4 $\times 10^4$  &   3.0 $\times 10^5$ \\ 
     \citet{Randall1992} & 1.5 $\times 10^4$  &   1.9 $\times 10^5$ \\ 
     \citet{Wyckoff1998} & 1.6 $\times 10^4$  &   3.3 $\times 10^5$ \\ 
     \citet{Newburn1989} & 1.8 $\times 10^4$  &  4.2 $\times 10^5$ \\ 
     \citet{Womack1994} & 2.5 $\times 10^4$ &   1.9 $\times 10^5$ \\ 
     \citet{Fink1991} & 2.8 $\times 10^4$  &   3.2 $\times 10^5$ \\ 
    \hline
    \end{tabular}
\label{tab:CN_Scale}
  \end{table}

A rare feature of C/2016 R2 is the weakness of the CN (0,0) B$^2\Sigma^+$ - X$^2\Sigma^+$ emission 
lines at 3880~\AA, usually one of the strongest emission features observed in optical spectra of 
comets. Here it is almost entirely dominated by the N$_2^+$ of first negative group (B$^2\Sigma_u^+$ - 
X$^2\Sigma_g^+$) (0,0) bandhead at 3914~\AA. 

In order to identify the CN emission lines, we used a theoretical fluorescence spectrum computed by interpolation at the right heliocentric velocity from a spectrum calculated by \citet{Zucconi1985}. 
This model was convolved by the instrument response function. The resulting synthetic CN model is shown in figure \ref{fig:Model}.

The CN lines are then identified and summed along the pixels of the spectrometer using the synthetic model as a mask. For each wavelength for which the CN spectrum is non-null, 
we identified a potential match. We used the high-resolution spectral atlas of cometary lines in 122P/de Vico \citep{Cochran2002} in order to identify potential emission lines due to other species than CN. In some cases, there is an overlap of several species, which both could produce the observed intensity. In particular, CH (0-0) is intense in C/2016 R2 (contrary to CN and C2), particularly around 3881~\AA~, around 3886~\AA~, and around 3892~\AA~. In the case where CN lines overlap with other species, the intensity was not measured, in order to avoid contamination. As these CN lines are remarkably weak compared to the intense overlap, these two lines were eliminated with no effect to the CN $g$ factor or the total flux. The selected lines were integrated in wavelength to compute their total flux $F_\text{tot}$ so as to draw an intensity profile on both sides of the comet's nucleus, which is shown in Fig. \ref{fig:avr}.

We then averaged the flux of each line for the observations of the three nights. The resulting averaged intensity profile is shown in black in Fig. \ref{fig:CN11}. The final intensity profile is a 1D profile starting from the nucleus of the comet and sweeping outwards along the coma ending at a distance of $\rho \sim 6.5 \times 10^3$~km. We first computed this profile separately for both side of the slit (with respect to the nucleus, located in the middle of the slit) and then averaged again these two 1D profiles into a single one. The total flux measured for CN over the entire slit and averaged over the three nights of observation was $2.1\times 10^{-15}$~erg.s$^{-1}$.cm${-^2}$. 

By using a $\chi^2$ test we were able to estimate the best fit of the Haser model to the observed intensity profile and determine the scalelengths of both the parent (HCN) and daughter (CN) species in the coma of C/2016 R2. The best value for the scalelength of HCN was determined to be $l_p = 1.3 \times 10^4$~km while for CN radical it was determined to be $l_d = 2.8 \times 10^5$~km, see Fig. \ref{fig:Chi} (values scaled to 1~au using an $r_h^{2}$ law). Our fit, as well as other fits using scalelengths from the literature, are shown after the convolution by the average seeing of the three nights in Fig. \ref{fig:CN11}. Table \ref{tab:CN_Scale} shows that our scalelengths values are in good agreement with the literature. It confirms that, despite a relatively small slit, we have a sufficient range of cometocentric distances to model the intensity distribution of a daughter product with a good accuracy. However, it should be noted that multiple pairs of scalelengths could also fit the data.

The branching ratio for CN produced by HCN is nearly equal to one \citep{Huebner1992}. The parent-species is most likely HCN as seen by the dust-poor composition of C/2016 R2 \citep{Opitom2019}.
Using the fluorescence factor $g = 3.52 \times 10^{-2}$ photons.s$^{-1}$.molecule$^{-1}$ at 1~au 
provided by \citet{Schleicher2010}, we estimate a production rate of $\sim 10^{25}$ mol.s$^{-1}$ 
if we assume the usual value of $v_d=1$~km.s$^{-1}$). This latter values is about three times the 
one published by \cite{Opitom2019}, based on a similar value of $v_d$.
The authors of \cite{Opitom2019} have since reviewed their computation about the overall CN flux measured
in the slit and an error was found. After correction, it agrees with our revised CN production rate. 
If we assume $v_d=0.5~km.s^{-1}$, as it seems closer to the reality
we obtain Q(CN)$\sim 5\times 10^{24}$ mol.s$^{-1}$. 

\subsection{N\texorpdfstring{$_2^+$}~ production rate\label{sec:n2}}

The intensity profiles can be used to get a determination of the scalelengths both for the parent molecule and the daughter species. N$_2^+$ is the daughter ion produced by the photoionization of N$_2$, so we apply the Haser model. We use the same parameters and methodology as those in our fit of the Haser profile to the CN intensity profile as it was shown to provide good results. From the synthetic fluorescence spectrum computed by \cite{Rousselot2022} we determined the line wavelengths of the N$_2^+$ (0,0) band and the corresponding lines in the UVES spectra. Here, we limit the identification process to the 3885.5~\AA ~to 3915.0~\AA ~interval in order to extract only the lines of the (0,0) band and to exclude the CN emission lines region. The Haser profile is then fit to our observations using the same methods as presented in Sec. \ref{sec:CN}. The resulting intensity profile is shown in Fig. \ref{fig:avr}.

\begin{figure}
	\includegraphics[width=\columnwidth]{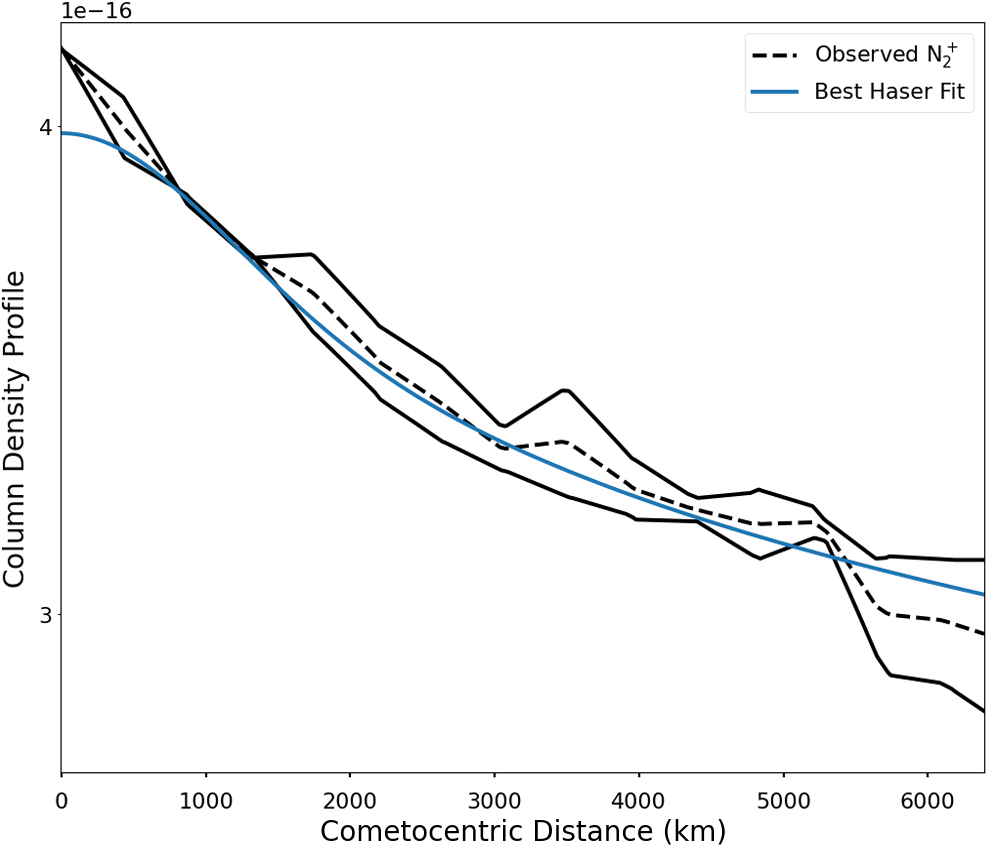}
    \caption{Best fit of the Haser model (blue) to the observed flux of N$_2^+$ (black). The observed N$_2^+$ flux is averaged from the observations of 11 Feb, 13 Feb, and 14 Feb (a single profile for both sides of the slit centred on the nucleus), then averaged again over cometocentric distance (dashed line).}
    \label{fig:N211}
\end{figure}

We explore this interval with the $\chi^2$ test and find new scalelengths of $l_p = 2.8 \times 10^6$~km and $l_d = 3.8 \times 10^6$~km (scaled to 1~au). This fit is shown in figure \ref{fig:N211} after being convolved by the average seeing of the three nights. These 
values are within the expected range estimated from the rate coefficients. However, at this scale, multiple pairs of scalelengths could be selected for N$_2^+$ with an almost equally good fit. The difference between the fit of different scalelength pairs is very small and the effect on the slope is almost negligible. We understand our intensity profile presents a certain degree of noise, and even the slightest alteration to our selection criteria would give a different best fit profile. 

These scalelengths can be used to compute the N$_2^+$ production rate. When we take into account the
branching ratio, only a small fraction of the N$_2$ molecules is ionized to N$_2^+$, most of them being 
photodissociated. \cite{Huebner1992} provide rate coefficients for the photodissociation and 
ionization of N$_2$.
From these data we computed that the branching ratio for creating N$_2^+$ ion is 0.34 for a quiet 
Sun (and 0.36 for an active Sun but with a total photodestruction rate being about twice). 
\cite{Wyckoff1976} also provide an estimate of both the photodissociation rate and 
photoionization rate, significantly lower than the one of \cite{Huebner1992} and a branching ratio
of 0.42. They admit, like \cite{Huebner1992}, that there is a large uncertainty on their values.
The photodissociation rate, especially, is rather uncertain because both of uncertainties on the
cross sections and the fact that a predissociation line has a similar wavelength to the solar
Lyman $\gamma$ line (highly variable with solar activity and that could be Doppler shifted).
We use an average of these two branching ratios, i.e. 0.38.

The fluorescence factor $g$ estimated by \cite{Rousselot2022} for the (0,0) 
band in the 3885.5~\AA\ to 3915.0~\AA\ interval is $5.41 \times 10^{-3}$~photon.s$^{-1}$.ion$^{-1}$ 
at $r_h = 2.76$~au. Using this value of $g$, the measured total integrated flux 
$F_\text{tot} = 1.0 \times 10^{-14}$~erg.s$^{-1}$.cm$^{-2}$, a branching ratio of 0.38,
and a velocity $v_d$ of 0.5~km.s$^{-1}$ at $r_h$ (see Section 3) we estimate a N$_2$ 
production rate of $\sim 1.0 \times 10^{28}$~molecules.s$^{-1}$.

Our result can be compared to other estimates based on the ratio with the CO production rate. A first result has been calculated by \citet{Wierzchos2018} as $Q(N_2) = (2.8 \pm 0.4) \times 10^{27}$ mol.s$^{-1}$ by determining the N$_2$ column density and production rate using the N$_2$/CO abundance ratio calculated from optical spectra and their own CO results, purposefully chosen so as to be as close as possible to \citet{McKay2019}, with Q(N$_2$) = (4.8 $\pm$ 1.1) $\times 10^{27}$ mol.s$^{-1}$. While not said explicitly in \citet{Biver2018}, their N$_2$ production rate can be inferred from their CO production rate and N$_2$/CO radio to be in the order of $Q(N_2) \sim 8.5 \times 10^{27}$ mol.s$^{-1}$. 
These production rates were all derived directly from fluxes by using the relative ratio of N$_2^+$ /CO$^+$, as:

\begin{equation}
\frac{N_2^+}{CO^+} = \frac{g_{CO^+}}{g_{N_2^+}}\frac{I_{N_2^+}}{I_{CO^+}} 
\end{equation}

\noindent where $I$ is the intensity of the bands. The relative ratio was determined using a $g$ factor of $7 \times 10^{-2}$ photons.s$^{-1}$.ion$^{-1}$ from \citet{Lutz1993} scaled to 1~au. We estimate these values with the updated $g$ factor from Rousselot et al. 2022, using $g = 4.90 \times 10^{-2}$~photons.s$^{-1}$.ion$^{-1}$ (for the whole band). Prior measurements of the N$_2$ production rate become Q(N$_2$) = 4.6 $\times 10^{27}$ mol.s$^{-1}$ for \citet{Wierzchos2018}, Q(N$_2$) = 8.0 $\times 10^{27}$ mol.s$^{-1}$ for \citet{McKay2019}, and Q(N$_2$) = 1.4 $\times 10^{28}$ mol.s$^{-1}$ for \citet{Biver2018}, providing the upper and lower values of the expected production rate. Our N$_2$ production rate, consequently, is right within the adjusted production rates found using relative ratios. It is, nevertheless, the first direct estimate of this production rate, independent from any assumption about the CO production rate. 

In our estimate of this production rate the main sources of uncertainties come from the determination
of $l_p$ and the radial expansion velocity of daughter products $v_d$. We also assume that the solar wind has little influence on the N$_2^+$ column density in the region close to the nucleus 
(the radial 
profile being then well described by the Haser model). Our model provides a value
for $l_p$ with the same order of magnitude as the estimates of N$_2$ destruction rates
provided by \cite{Huebner1992} and \cite{Wyckoff1976}, but the difference between these two estimates
is large (a factor of three). A lower value of $l_p$ would reduce the N$_2^+$ production rate 
roughly in the same proportion, leading to a significantly different value of the one computed 
from the N$_2^+$/CO$^+$ ratio and CO production rate. The relationship from these two parameters
is based on a similar ionization rate for N$_2$ and CO.
It must also be kept in mind that photodestruction of N$_2$ is 
strongly dependant on solar activity. Our estimate of $l_p$ corresponds to a low solar activity
and could be significantly different for comets observed during maxima of solar activity.
Errors also arise from our estimation of the total flux $F_\text{tot}$: Our selection criteria on the emission lines can be more restrictive or permissive, depending on whether we identify a match where our model is non-null, or try to restrict noise by setting a positive lower limit. However, this only has an effect of $\pm 0.1$ erg.s$^{-1}$.cm$^{-2}$.

Another important result from our work concerns the variability
of N$_2$ production rate between observation dates. The production rate given above corresponds 
to the average of our three 
observing nights but, as shown Figure \ref{fig:avr}, the observed intensity significantly changes 
from one night to another. If we compute N$_2$ production rates separately for these three nights
with the same parameters mentioned above we find, respectively for 11, 12 and 13 Feb.: $1.0 \times 
10^{28}, 8.0 \times 10^{27}$, and $1.2\times 10^{28}$.
Such rapid changes in the N$_2$ line intensities are not an instrumental effect because CN profiles 
do not reveal any similar change ($F_\text{tot} = 2.31, 2.01, 2.01 \times 10^{-15}$ 
erg/s/cm$^2$ for $Q$(CN) = 5.4, 4.7, 4.7 $\times 10^{24}$ mol.s$^{-1}$ on Feb 11, 13, and 14, 
respectively). The origin of such rapid changes is unknown but likely related to inhomogeneities of the nucleus, thought it may be due to changes in the solar wind, as we do not see such such rapid changes for neutral species.

\section{Summary and Conclusions}

We investigated the CO-dominated and water-poor comet C/2016 R2 (PanSTARRS) using the Haser model in order to determine the N$_2$ production rate. The intensity of the N$_2^+$ lines is truly incomparable to what has been seen in comets so far, confirming the rarity of this type of comet. We determined the following characteristics:

\begin{itemize}
\item[$-$] Parent (HCN) and daughter (CN) scalelengths of $l_p = 1.3 \times 10^4$~km and $l_d = 2.8 \times 10^5$~km respectively (scaled to 1 au using $r_h^2$), consistent with the literature;
\item[$-$] A CN production rate of $Q$(CN)$\sim 5\times10^{24}$~molecules.s$^{-1}$, 
lower than usually observed in cometary spectra at similar heliocentric distances but consistent with other estimates for this comet;
\item[$-$] Parent (N$_2$) and daughter (N$_2^+$) scalelengths of $l_p = 2.8 \times 10^6$~km and 
$l_d = 3.8 \times 10^6$~km respectively (for 1~au, when using a scaling with $r_h^2$). ;
\item[$-$] A N$_2$ production rate of $Q$(N$_2$) $\sim 10^{28}$~molecules.s$^{-1}$,
exceptionally high, within the values estimated by other teams from the ratio N$_2^+$/CO$^+$;
\item[$-$] When compared to a CO production rate of Q(CO) $\sim$ 1.1 × 10$^{29}$ molecules.s$^{-1}$, we find a N$_2$/CO ratio of 0.09, which is consistent with observed intensity ratios. This is the highest of such ratios observed for any comet so far;
\item[$-$] Some large variations of N$_2$ production rates over the course of the three observing nights, the 
N$_2$ production rate given above being the average of these values. The production rate computed
with the same parameters varies between $8.0 \times 10^{27}$ and $1.2 \times 10^{28}$ molecules.s$^{-1}$.
\end{itemize}

Table~\ref{tab:table2} summarises these results.

\begin{table}
	\centering
	\caption{Parent and daughter scalelengths calculated for CN and N$_2^+$ (at 1~au, scaled with
	$R_h^2$) and the resulting production rates for their parent species.}
	\begin{tabular}{lccr} 
		\hline
		 & $l_p (\times 10^3$ km) & $l_d (\times 10^3$ km) & Q (molecules.s$^{-1}$) \\
		\hline
		CN & 13 & 280 & $5 \times 10^{24}$ \\
		N$_2^+$ & 2800 & 3800 & $1 \times 10^{28}$\\
		\hline
	\end{tabular}
	\label{tab:table2}
\end{table}

Large uncertainties still remain, and as it is so far the only comet of this kind to ever be observed with modern equipment, we are lacking a point of comparison. A detailed observation of C/2016 R2 at high heliocentric distances could still be possible, as CO would continue to sublimate under 40 au,  
and would provide further detail of the nucleus while inactive, in order to create a detailed thermal and structural model of this peculiar comet. 

\section*{Data availability}

No new data were generated or analysed in support of this research.




\bibliographystyle{mnras}
\bibliography{references} 








\bsp	
\label{lastpage}
\end{document}